\documentclass[preprint]{scrartcl}

\usepackage{graphicx}
\usepackage{amsmath}
\usepackage{amsfonts}
\usepackage{bbold}
\usepackage{braket}
\usepackage{siunitx}
\usepackage[breakable]{tcolorbox}

\newtcolorbox{highlighted}{colback=yellow,coltext=red,breakable}

\usepackage{subfig}

\newcommand{\sa}{s_{\alpha}}
\newcommand{\ef}[1]{\epsilon_{{\textnormal{eff}}#1}}
\newcommand{\psqd}[1]{P_{{\textnormal{SQD}}#1}}
\newcommand{\pmnp}{P_{\textnormal{MNP}}}

\date{}

\begin{document}

\title{Generation of maximally entangled states in hybrid two quantum dots mediated by a spherical metal nanoparticle driven by external laser field}

\author{Kostas Blekos$^a$ \and Maria-Eftaxia Stasinou$^a$ \and Andreas F. Terzis$^a$ \and Emmanuel Paspalakis$^b$}

\maketitle

\begin{abstract}
We theoretically study the generation of quantum correlations 
in a hybrid system composed by two interacting semiconductor quantum dots mediated by a metal nanoparticle 
and coupled to an external laser field.
Interactions present in the hybrid system are treated using a semiclassical approximation except for the direct dipole-dipole interaction. 
We report the entanglement of formation, which gives information about entanglement quantum correlations, for continuous wave and pulsed driving applied fields.
We have found that for proper values of the physical and geometrical parameters of the hybrid system the applied field
can be tuned producing quantum correlations of significant value 
so that they can be useful in quantum information and computation processes.\footnote{$^a$Department of Physics, School of Natural
Sciences, University of Patras, Patras 265 04, Greece\\
$^b$Materials Science Department, School of Natural
Sciences, University of Patras, Patras 265 04, Greece}
\footnote{Submited September 12, 2015, updated: March 17, 2017}
\end{abstract}

\section{Introduction}

Generation of well-controlled quantum correlations is the main research of many scientists working in the relatively new field of quantum information and quantum processing~\cite{nielsen}.  
The most thoroughly studied quantum correlation is the entanglement. 
This striking feature has been mainly quantified by means of the concurrence and the entanglement of formation (EoF)~\cite{Wootters}.
A prototype system, the simplest composite system which can display quantum entanglement, is a two-qubit system. 
Pairs of semiconductor quantum dots (SQDs) where each one is characterized by two-level states 
(ground state and excitonic excited state) 
are ideal experimentally realistic nanostructures. 
Moreover, by introducing a plasmonic nanostructure between the two SQDs, 
the local electromagnetic field can be altered in a somehow desirable way 
and strong quantum correlations can be obtained
\cite{plasm1,plasm2,plasm4,He12a,plasm3,susa,Bryant13a,kinezoiOIP,Li13a,angelakis14a,susa14,Racknor14a,hou14a,Nerkararyan15,Hughes15a,otten_entanglement_2015,{schindel_study_2015},{yang_analytical_2015},{nugroho_tailoring_2015},{terzis_nonlinear_2016},otten_origins_2016,iliopoulos_two-qubit_2016,sadeghi_2017}.

In the above area of studies, Artuso and Bryant~\cite{Bryant13a} studied a system composed by two SQDs 
(modeled as two-level systems) 
mediated by a spherical metal nanoparticle (MNP). 
They gave emphasis to the interaction of the system with an external electromagnetic field 
and presented the population dynamics of the two-qubit system for several cases.
Previously, He and Zhu~\cite{He12a} had studied the transient population dynamics and the entanglement dynamics of the same system in the absence of an external electromagnetic field. 
Also, quite recently, Nerkararyan and Bozhevolnyi~\cite{Nerkararyan15} have investigated 
how the relaxation dynamics in a system of two quantum dipole emitters 
(modeled as three-level systems) 
coupled to a spherical MNP influences the entanglement dynamics.

The externally-driven coupled SQD --- spherical MNP --- SQD system~\cite{Bryant13a} 
is a direct extension of the externally-driven coupled SQD --- spherical MNP system, 
which is a basic system of quantum plasmonics~\cite{Tame13a}. 
In the externally-driven coupled SQD --- spherical MNP system, 
the optical properties and the controlled population dynamics has been extensively studied 
\cite{Wang06a,Sadeghi09a,Ridolfo10a,Artuso10a,Malyshev11a,Hatef11a,Kosionis13a,Paspalakis13a,Hatef13b,Malyshev13a,Tasgin13a,Sadeghi13b,Anton13a,Li14a,sadeghi_ultrafast_2015}. 
In this work, we return to the externally-driven coupled SQD --- spherical MNP --- SQD system~\cite{Bryant13a} 
and study in detail the entanglement between the two SQDs. 
We quantify the entanglement with the EoF 
and present results for both transient entanglement dynamics and the steady state value 
of entanglement for various system parameters.

\begin{figure}[htp]
	\centering
	\includegraphics{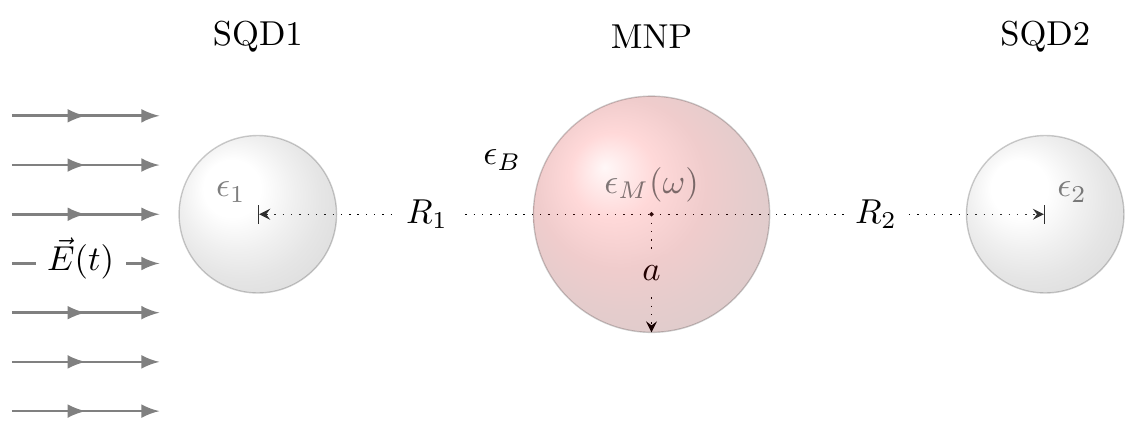}
	\caption{SQD --- MNP --- SQD geometry}\label{fig:sqdgeometry}
\end{figure}

This paper is organized as follows. 
In sect. 2, 
 we present the theoretical model for the hybrid nanostructure in the presence of an applied external electromagnetic field 
 and introduce a semi-classical scheme 
 to describe the coupling of the pair of qubits directly to the applied field and through the MNP particle. While the direct SQD-SQD coupling is modeled quantum mechanically. 
Next, in sect. 3 
 we present the results for the time evolution of EoF and steady state results for EoF 
 as a function of the various characteristic parameters of the hybrid system for both cw and pulsed fields. 
Finally, in sect. 4 
we summarize our findings.

\section{Theory}

We consider a system as previously described in~\cite{Bryant13a}, 
 where two identical SQDs (labeled by index $i (=1,2)$) 
 mediated by a spherical metal nanoparticle of radius $\alpha$ 
 are subject to a continuous sinusoidal wave (part 1) and pulsed laser (part 2) electric field (fig~\ref{fig:sqdgeometry}).
The SQDs are at a center-to-center distance of $R_i$ from the MNP 
 and the driving field of the applied cw is characterized by its magnitude $E_0$ and angular frequency $\omega$. 
 For the pulsed laser field case, we assume a hyperbolic secant envelope, 
 $f(t), E = E_0f(t)\cos(\omega t)$ with $f(t) = \textnormal{sech}\left((t-t_0)/t_p\right)$, 
 though the exact envelop shape is not of particular importance.
The SQDs are directly coupled to the applied field, 
 feel the electric field produced by the MNP and interact with each other.

We only consider two levels for the SQDs having exciton energies $\hbar\omega_i$, 
transition electric dipole moments $\mu_i$ 
and dielectric constants $\epsilon_i$. $\epsilon_B$ is the dielectric constant for the material in which the entire system is embedded.

The Hamiltonian for the system is
\begin{equation}
	H_{\textnormal total} = H_0 + H_{\textnormal SQD} + H_{\textnormal int}
\end{equation}
where 
$H_0$ the unperturbed Hamiltonian,
$H_{\textnormal SQD}$ describes the SQD --- MNP interaction and
$H_{\textnormal int}$ is the direct coupling between the SQDs.

We write $H_0$ as 
$H_0 = \hbar\omega_i\hat{\alpha}_i^\dagger \alpha_i$, 
where repeated indexes imply summation over the indexes
and $\hat{\alpha}^{(\dagger)}_i$  are the exciton annihilation (creation) operators for each SQD.\@

\subsection{SQD --- MNP interaction}
We calculate $H_{\textnormal SQD}$ semi-classically, so
$H_{\textnormal SQD} = -\mu_i (E_{SQDi}^*\hat{\alpha}_i^\dagger+ E_{SQDi}\hat{\alpha}_i)$.
$E_{SQDi}$ is the electric field at the center of SQD$i$. 

We write the electric field of the SQDs in terms of 
the screening factor ($\ef{i}=\frac{2\epsilon_B+\epsilon_i}{3\epsilon_B}$) 
and the induced field produced by the polarization of the MNP ($E_{M,i}$):
\[E_{SQDi}=\frac{1}{\ef{i}}(E(t)+E_{M,i}) \, ,\]
where the induced field and the polarization of the MNP are in turn~\cite{Bryant13a}
\begin{align*}
	E_{M,i} &= \frac{1}{4\pi e_B}\frac{\sa \pmnp}{R_i^3} \\
	\pmnp &= 4\pi\gamma a^3 e_B\left( E(t) + \frac{\sa}{4\pi e_B}\sum_i \frac{\psqd{i}}{R_i^3\ef{i}} \right)
\end{align*}

The polarization $\psqd{i}$ of each SQD is found from the density matrix elements,
for the allowed transitions:
\begin{align*}
	\frac{\psqd{i}}{\mu_i} &= \ket{1} \rho \bra{\textnormal{SQD}i_g} + \ket{\textnormal{SQD}i_g}\rho\bra{1} 
		+ \ket{4} \rho \bra{\textnormal{SQD}i_e} + \ket{\textnormal{SQD}i_e} \rho \bra{4}\\
\frac{\psqd{1}}{\mu_1} 	&= \rho_{1,2} + \rho_{2,1}+ \rho_{4,3} + \rho_{3,4} \\
\frac{\psqd{2}}{\mu_2} 	&= \rho_{1,3} + \rho_{3,1}+ \rho_{4,2} + \rho_{2,4} \\
\end{align*}
or, compactly, $\frac{\psqd{i}}{\mu_i}  =\tilde{\rho}_{pi} + {\tilde{\rho}_{pi}}^*$,
defining $ \tilde{\rho}_{p1} = \rho_{1,2} + \rho_{3,4}$ and $\tilde{\rho}_{p2} = \rho_{1,3} + \rho_{2,4}$.
Our basis is composed of four states, 
named 
  $\ket{1}=\ket{g_{1}g_{2}}$, 
  $\ket{2}=\ket{g_{1}e_{2}}$, 
  $\ket{3}=\ket{e_{1}g_{2}}$ and 
  $\ket{4}=\ket{e_{1}e_{2}}$, 
where $g_{i}$ and $e_{i}$ label the ground and the excited states of the $i$-qubit, respectively. 

To simplify the expression for $\pmnp$ we introduce the parameters $G_i, \Omega_i$ and $F$:
\begin{align*}
G_i&=\frac{\gamma\alpha^3s_{\alpha}^2\mu_i^2}{4\pi\epsilon_B\ef{i}^2R_i^6\hbar} \\
\Omega_i&=\frac{E_0\mu_i}{2\hbar\ef{i}}(1+\frac{\gamma\alpha^3s_{\alpha}}{R_i^3}) \\
F&=\frac{1}{4\pi\epsilon_B}\frac{\gamma\alpha^3s_{\alpha}^2\mu_1\mu_2}{\hbar\ef{1}\ef{2}R_1^3R_2^3}
\end{align*}
so we can now write:
\begin{align}
	E_{{\textnormal SQD}i} &= \frac{G_i\hbar}{\mu_i^2}\psqd{i} + \frac{F\hbar}{\mu_i\mu_{\bar{\imath}}}\psqd{\bar{\imath}} 
	+ 2f(t)\frac{\Omega_i\hbar}{\mu_i}\cos{\omega t}\nonumber \\
	H_{\textnormal SQD} &= -\left(\frac{G_i\hbar}{\mu_i}\psqd{i} + \frac{F\hbar}{\mu_{\bar{\imath}}}\psqd{\bar{\imath}} 
	+ 2f(t)\Omega_i\hbar\cos{\omega t}\right)\hat{\alpha}_i + \textnormal{h.c}
\end{align}

\subsection{SQD direct coupling}
For the direct coupling the Hamiltonian can be written~\cite{{Bryant13a},{lehmberg}}
\begin{equation}
	H_{\textnormal int} = \hbar\delta
		(\hat{\alpha}_1 + \hat{\alpha}_1^{\dagger})
		(\hat{\alpha}_2 + \hat{\alpha}_2^{\dagger})\, ,
\end{equation}
where $\delta$ is the interaction energy between the SQDs.
In the model under study, all interaction fields are parallel to each other so we can write~\cite{{susa},{Bryant13a},{lehmberg}}
\begin{equation*}
	\delta(\omega,R_{1\leftrightarrow 2}) = -\gamma_{\textnormal em}\cdot\frac{3}{2}\left( \frac{\cos{\zeta}}{\zeta^3} + \frac{\sin{\zeta}}{\zeta^2} \right)
\end{equation*}
where $\zeta = \frac{\omega}{c}(R_1 + R_2)$ 
and $\gamma_{\textnormal em}$ is the decay rate that characterizes the interaction.

In case we apply a semiclassical approximation for the direct coupling of the two quantum dots 
we report very different results 
that do not fully reveal the quantum mechanical nature of our system. 
However we have observed that the same 
(fully quantum treatment) 
is not necessary for the other interaction terms.

\subsection{Rotating Wave Approximation and master equation}
The final Hamiltonian is
\begin{align*}
	H= & \hbar\omega_i\hat{\alpha}_i^{\dagger}\alpha_i \\
	   & -\hbar\alpha_i \cdot\left(2\Omega_i f(t)\cos{\omega t} + 
	  \frac{\psqd{i}}{\mu_i} G_i + \frac{\psqd{\bar{\imath}}}{\mu_{\bar{\imath}}} F \right)\\
	& -\hbar\alpha_i^{\dagger} \cdot\left(\textnormal{h.c.}\right) \\
	& +\hbar\delta (\hat{\alpha}_1 + \hat{\alpha}_1^{\dagger}) (\hat{\alpha}_2 + \hat{\alpha}_2^{\dagger})\, ,
\end{align*}

We transform the Hamiltonian to a different basis:
\[
	\ket{1} \to e^{i\omega t}\ket{1} ,
	\ket{2} \to \ket{2},
	\ket{3} \to \ket{3},
	\ket{4} \to e^{-i\omega t}\ket{4}.
	\]
In this basis $\psqd{i}$ is
\begin{align*}
	\frac{\psqd{i}}{\mu_i} &=  e^{i\omega t} \tilde{\rho}_{\textnormal{p}i} + e^{-i\omega t} \tilde{\rho}^*_{pi}
\end{align*}
and the full Hamiltonian is, after a rotating wave approximation:
	\begin{align}
		H_{\textnormal total} =& \hbar(\omega_i-\omega)\hat{\alpha}_i^{\dagger}\alpha_i + \hbar\omega I_4 \nonumber\\
		   &-\hbar\,\alpha_i\cdot\left(\Omega_i f(t) + 
		   G_i\tilde{\rho}_{pi} + F\tilde{\rho}_{p\bar{\imath}}\right) \nonumber\\
		& -\hbar\,\alpha_i^\dagger\cdot\left( \textnormal{h.c}\right)  \nonumber\\
		& +\hbar\delta (\hat{\alpha}_1 \cdot \hat{\alpha}_2^{\dagger}) + \textnormal{h.c}  \nonumber\\
		=&-\hbar\begin{pmatrix}
	   -\omega & H_1& H_2 & 0 \\
	   H_1^{\star} & -\omega_1 & 0 & H_2 \\
	   {H_2}^{\star} & 0 & -\omega_2 & H_1 \\
	   0 & {H_2}^{\star}  & {H_1}^{\star} & \omega-(\omega_1+\omega_2)
	   \end{pmatrix}
	   +
	   \hbar\begin{pmatrix}
		   0 & 0 & 0 & 0\\
		   0 & 0 & \delta & 0\\
		   0 & \delta & 0 & 0\\
		   0 & 0 & 0 & 0
	   \end{pmatrix}
	\end{align}
where $H_1 = \Omega_1 f(t) + G_1\tilde{\rho}_{p1} + F\tilde{\rho}_{p2}$ and
$H_2 = \Omega_2 f(t) + G_2\tilde{\rho}_{p2} + F\tilde{\rho}_{p1}$

The master equation for the system is
\begin{equation}
\frac{d\rho}{dt} = \frac{i}{\hbar}[\rho,H] - \Gamma({\rho}) \, ,
\label{eq:master}
\end{equation}
where $\rho(t)$ is the density matrix of the system of the SQDs.

$\Gamma(\rho)$ is the relaxation matrix that describes the dissipative processes.
We use the approximation by~\cite{Bryant13a}, treating the SQDs as non interacting.
We calculate $\Gamma$ in terms of the density and relaxation matrices of each single SQD,
The final relaxation matrix is given by
\begin{align*} \Gamma(\rho) =& \left(\frac{1}{T_1}(\mathbb{1}\otimes\sigma_x)\circ\rho\right) +
                  \left(\frac{1}{T_2}(\sigma_x\otimes\mathbb{1})\circ\rho\right) +\\
	      &   \left(-\frac{1}{\tau_2}(\sigma_z\otimes{[\rho]}_{34}) \right) +
		\left(-\frac{1}{\tau_1}({[\rho]}_{24}\otimes\sigma_z) \right) \, ,
\end{align*}
where $\mathbb{1} = \begin{pmatrix}1 & 1\\ 1 & 1 \end{pmatrix}$, 
$\sigma_i$ the Pauli matrices, 
$\circ$ signifies pairwise product between matrices 
and ${[\rho]}_{ab} = \begin{pmatrix} \rho_{aa} & \rho_{ab}\\ \rho_{ba} & \rho_{bb} \end{pmatrix}$.
The parameters $T_i$ and $\tau_i$ are relaxation times characteristic of the system: 
$\tau_i$ pertains to spontaneous decay of the exciton state,
while $T_i$ pertains to decoherence by phonon interactions inside the SQD\@.

\subsection{Entanglement measure}
Finally, we present the measure of quantum correlations which is used in this work, i.e.\ the EoF. 
In order to find the EoF~\cite{Wootters} 
we compute the matrix $R = \rho(t)(\sigma Ay \otimes\sigma By ){\rho(t)}^*(\sigma Ay \otimes\sigma By )$, 
where 
$\rho(t)$ is the solution of the previous master equation 
and $\sigma iy$, with $i = A,B$
are the Pauli matrices of the two SQDs respectively. 
By diagonalising the matrix $R$, we
compute its eigenvalues $\lambda_j$, with $j = 1, 2, 3, 4$, 
and then we calculate the concurrence, 
which is $C = \max[0,\sqrt{\lambda_1}-\sqrt{\lambda_2}-\sqrt{\lambda_3}-\sqrt{\lambda_4}]$ 
for $\lambda_1 > \lambda_2 > \lambda_3 > \lambda_4$. 
Given that, we can compute the EoF expressed as
\begin{equation}
	\textnormal{EoF}(\rho) = h(\frac{1+\sqrt{1-C^2 (\rho)}}{2}) \, ,
\label{eq7}
\end{equation}
where $h (x)=-x\log_2 x-(1-x)\log_2 (1-x)$ is the binary entropy function. 
The EoF takes values from 0 (no entanglement for the system) to 1 (maximum entanglement).

\subsection{Parameters for the numerical calculations}
We summarize the  parameters $G_i,F,\gamma, \Omega_i,\delta$
\begin{align*}
G_i&=\frac{\gamma\alpha^3s_{\alpha}^2\mu_i^2}{4\pi\epsilon_B\ef{i}^2R_i^6\hbar} \\
\Omega_i&=\frac{E_0\mu_i}{2\hbar\ef{i}}(1+\frac{\gamma\alpha^3s_{\alpha}}{R_i^3}) \\
F&=\frac{1}{4\pi\epsilon_B}\frac{\gamma\alpha^3s_{\alpha}^2\mu_1\mu_2}{\hbar\ef{1}\ef{2}R_1^3R_2^3}\\
\delta &= -\gamma_{\textnormal em}\cdot\frac{6}{4}\left( \frac{\cos{\zeta}}{\zeta^3} + \frac{\sin{\zeta}}{\zeta^2} \right)\\
\zeta &= \frac{\omega}{c}(R_1 + R_2) \\
\gamma &=\frac{\epsilon_M(\omega)-\epsilon_B}{2\epsilon_B+\epsilon_M(\omega)}\, ,
\end{align*}
with $\epsilon_M(\omega) $ being the dielectric function of the MNP,\@
which we consider as that of gold, found experimentally~\cite{christy}.
We set $\epsilon_B=\epsilon_0$ and $\epsilon_1=\epsilon_2=6\epsilon_0$
and use $\ef{i}=\frac{2\epsilon_B+\epsilon_i}{3\epsilon_B}$.
We take $\sa = 2$ as all interaction fields are parallel to each other.

\begin{figure}[htp]
\centering
\includegraphics[scale=0.7]{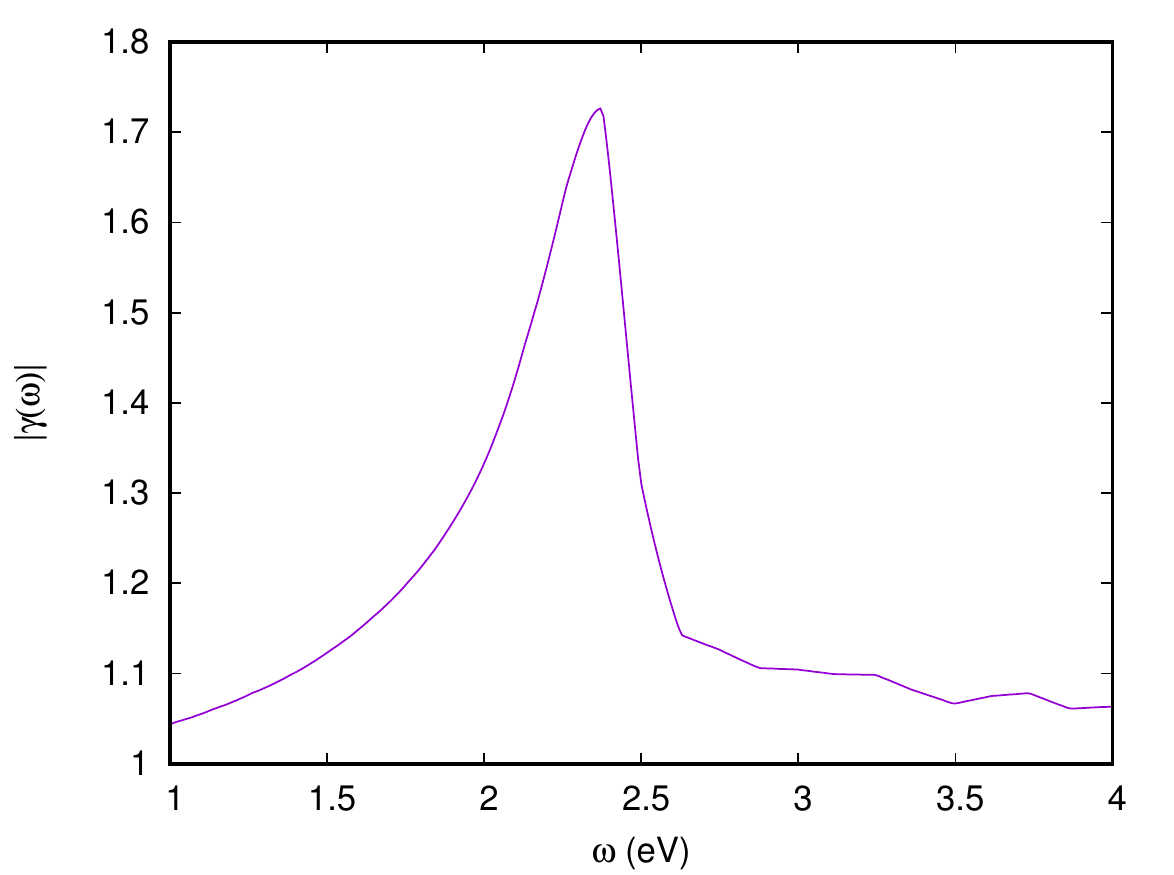}
	\caption{Experimentally found Dielectric response of gold~\cite{christy}}\label{fig:gamma}
\end{figure}

We set $\omega_1 = \omega_2 = \SI{2.5}{eV}$, which is near the plasmon peak for gold (fig.\ref{fig:gamma}).

For the radius of the MNP, $\alpha$, we use values in the range \SIrange{0.1}{8}{nm}.
We note that not all values in this range are practical but we use them
nevertheless for completeness of the numerical analysis.

Also, we take $T_{1} = T_{2} = \SI{0.3}{ns}$ and $\tau_{1} = \tau_{2} = \SI{0.8}{ns}$.
For $\gamma_{\textnormal em}$ we use the geometric mean of the spontaneous decay rates of each SQD, 
$\gamma_{\textnormal em} = \sqrt{1/T_1T_2}$

The characteristic times of the pulse are $t_0=\SI{22.5}{ps}$ and $t_p=\SI{3}{ps}$.

\section{Results}

In order to calculate the quantum correlations, 
we solve the set of differential equations of the density matrix elements of eq.\ref{eq:master}.
We calculate the EoF which is a measure of the quantum correlations related to entanglement. In the present study our initial state is the most natural one, $\ket{1}=\ket{g_{1}g_{2}}$,, where both qubits are in the ground state. 
An external laser field is applied to the system. 
The intensity of the applied laser field is proportional to the square magnitude of $E_{0}$. 

\begin{figure}[htp]
\centering
\includegraphics[scale=0.8]{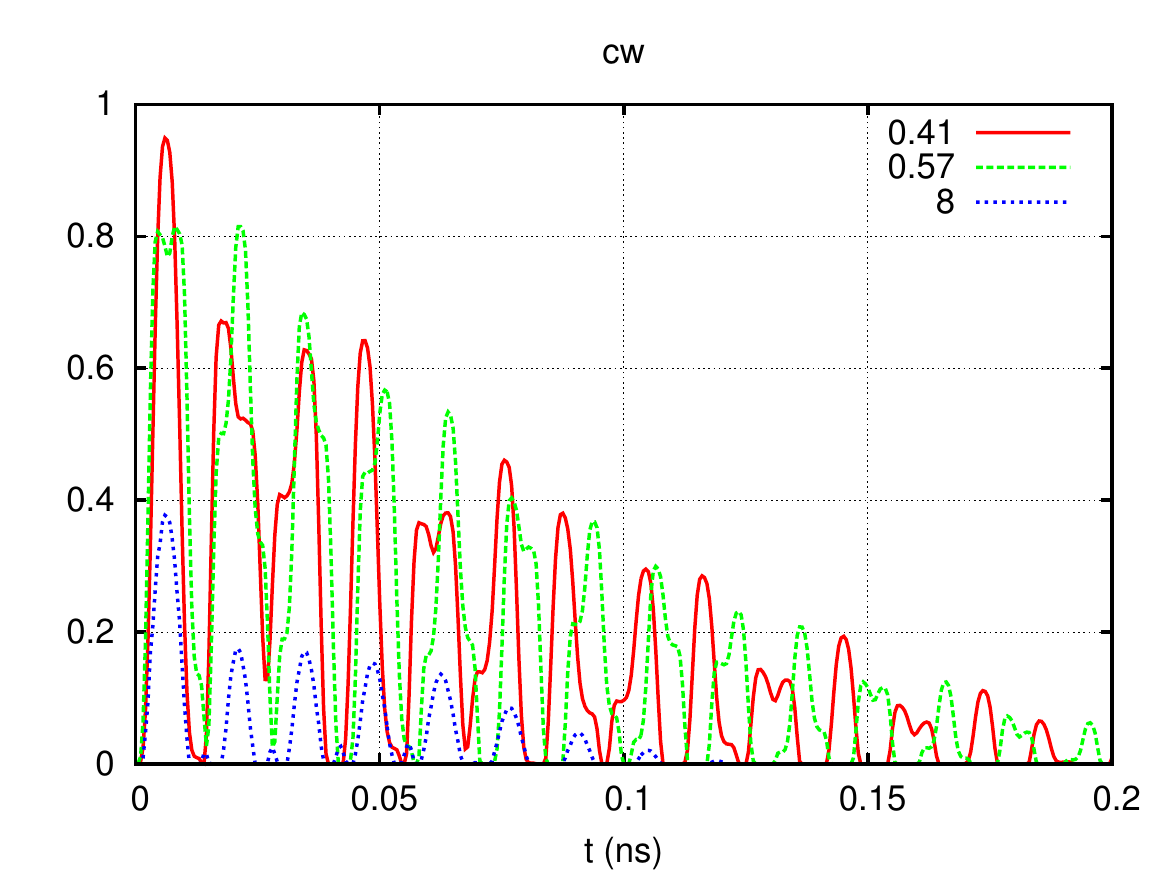}
\includegraphics[scale=0.8]{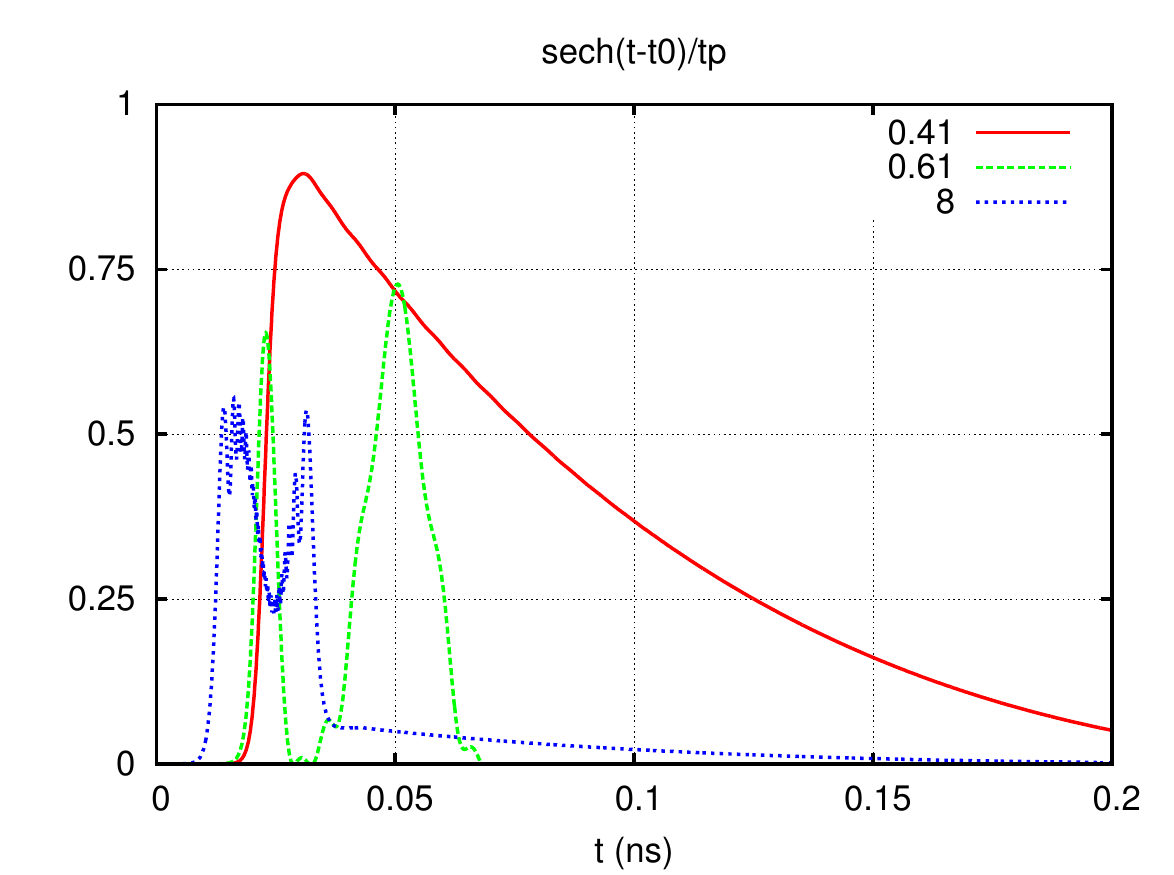}
\caption{Plots of EoF as a function of time for a hybrid system on resonance.
	Radius of MNP is 2 nm, the dipole moment of each SQD is \SI{2.20}{enm} and
	the distance of each SQD from the MNP is \SI{9}{nm}.
	Various plots are depicted 
	for characteristic values for the intensity of the field ($\SI{1e6}{V/m}$).
}\label{fig:eoftime}
\end{figure}

In fig.~\ref{fig:eoftime} we plot the time evolution of EoF 
for representative values of the geometric parameters of the hybrid system 
($\alpha = \SI{2}{nm}$, $\mu_{1}=\mu_{2}=\SI{2.20}{enm}$ and $R_{1}=R_{2}=9$ nm). 
We apply a cw and pulsed laser field 
(of $t_0=\SI{22.5}{ps}$ and $t_p=\SI{3}{ps}$) 
with angular frequency of the field equals the excitonic transition eigenfrequency (resonance condition) 
and various characteristic intensities (for comparison and consistency see next figure, fig.\ref{fig:eof-intensity}). 
First for the cw field we observe that 
EoF reaches a maximum value strongly depended on the magnitude of the applied field. 
For a value of 0.41 we get the maximum value (approximately 0.9). 
In all cases studied we have found that the maximum value is achieved at an early time
and then the EoF performs an oscillatory decline, which is in accord o the characteristic relaxation times. 
Similarly, for the pulsed field 
the EoF reaches a maximum value of approximately 0.8 at the same intensity as in the cw case, and then smoothly declines. 
We emphasize here that we have observed this pattern in all cases that we investigated, 
for the same parameter set, 
the cw field achieves slightly higher EoF values 
that oscillate chaotically to lower values.
Actually, two facts make the  pulsed field 
more practical for use in quantum computation processes. 
First 
the gradual declination of EoF 
and second the observation that the maximum value of EoF can be related to the shape of the pulse 
(for example the peak position).

Then, in fig.\ref{fig:eof-intensity}, we investigate the dependence of maximum EoF as a function of the intensity of the applied field for various angular frequencies of the field (on resonance and on off-resonance conditions). These are very informative plots and clearly reveal the importance of the intensity of the externally applied laser field on achieving high values of entanglement in a robust manner. A very significant observation is that in most cases investigated, above a characteristic value of the intensity the maximum EoF achieved has the same (relatively) high value. Moreover, in the off-resonance case there is a strong dependence on the position of the two SQDs (see the plot in the right column and middle row). 

Then, we systematically study the dependence of EoF of the hybrid system on the various parameters of the system. In all cases studied we apply a cw and pulsed laser (of $t_0=\SI{22.5}{ps}$ and $t_p=\SI{3}{ps}$) with intensity, found under systematic numerical investigation,  that maximizes the observed EoF. 
Hence, we mainly study maximum EoF as a function of the hybrid system geometrical parameters, 
i.e.\ size of MNP, dipole moments and the position of the SQDs, 
and the external parameters, 
i.e.\ the characteristic parameters of the external fields.

First, in fig.~\ref{fig:dens_alfa-miu-max-e0} we depict the density diagram of the maximum achieved EoF 
 as a function of dipole moments of the two identical SQDs,  $\mu_{1}=\mu_{2}$  
 and the radius of the MNP $\alpha$ 
 for $R_{1}=R_{2}=9$ nm.
From this diagram it appears that we have two distinct regions. 
One where we can find very high EoF values
and a region of significantly lower EoF values.
The observed behavior is in accordance to the one reported in top fig. 3 of ref.~\cite{Bryant13a}. 
The region of high EoF corresponds to the EXIT and the transition chaotic region of the phase diagram there~\cite{Bryant13a}. 
The transition suppression and the bistability region shows lower EoF. 
In the next figure (Fig.~\ref{fig:dens_alfa-miu-max-e0vfinal}) we investigate the final EoF for the pulsed case, as the final EoF of the cw filed is always zero.
We report that the shorter distance of $R_{1}=R_{2}=9$ nm case gives higher values of the EoF compared to the higher distance of $R_{1}=R_{2}=13$ nm case.
 
Finally, we have found a rather strong dependence of angular frequency of the laser field. In Fig.~\ref{fig:dens_Dipole-detuning} 
 we report for fixed radius of the MNP ($\alpha=1nm$) the maximum EoF as a function of the dipole moment of the SQDs and the detuning of the applied field and found a very strong dependence.
Actually, fig.~\ref{fig:dens_Dipole-detuning} is highly asymmetric, 
showing significant values of EoF only for negative detuning 
 or for very close to resonance driving fields. We have found that for positive detuning, i.e. for incident photons of higher energy than the exciton energy, the EoF shows low values, strongly depending on the detuning

\section{Summary}
We have theoretically investigated the entanglement 
 created in a hybrid nanostructure molecule consisting of two interacting SQDs coupled to a centrally located MNP 
 driven by an external applied cw and pulsed laser field. 
The SQDs excitons are modeled quantum mechanically 
 as two level systems 
 and the MNP classically as a spherical MNP\@. 
Also, all interactions in the system are treated using a semiclassical approximation, except the direct SQDs coupling. 
The dynamics of the hybrid system is studied 
 by means of the master equation of the density matrix of the two SQDs in the presence of the MNP\@. 
We use the density matrix elements and calculate the EoF numerically.
In general, we have found that for proper values 
 of the dipole moments of SQDs, 
 the size of MNP 
 and the distance of the SQDs from the MNP, 
 as well as for the intensity of the applied, cw or pulsed field, 
 we can produce steady state quantum correlations of significant value, 
 and maximally entangled states
 so that they can be useful in quantum information and computation processes.

{}

\begin{figure}[htp]
\centering
\includegraphics[scale=0.45]{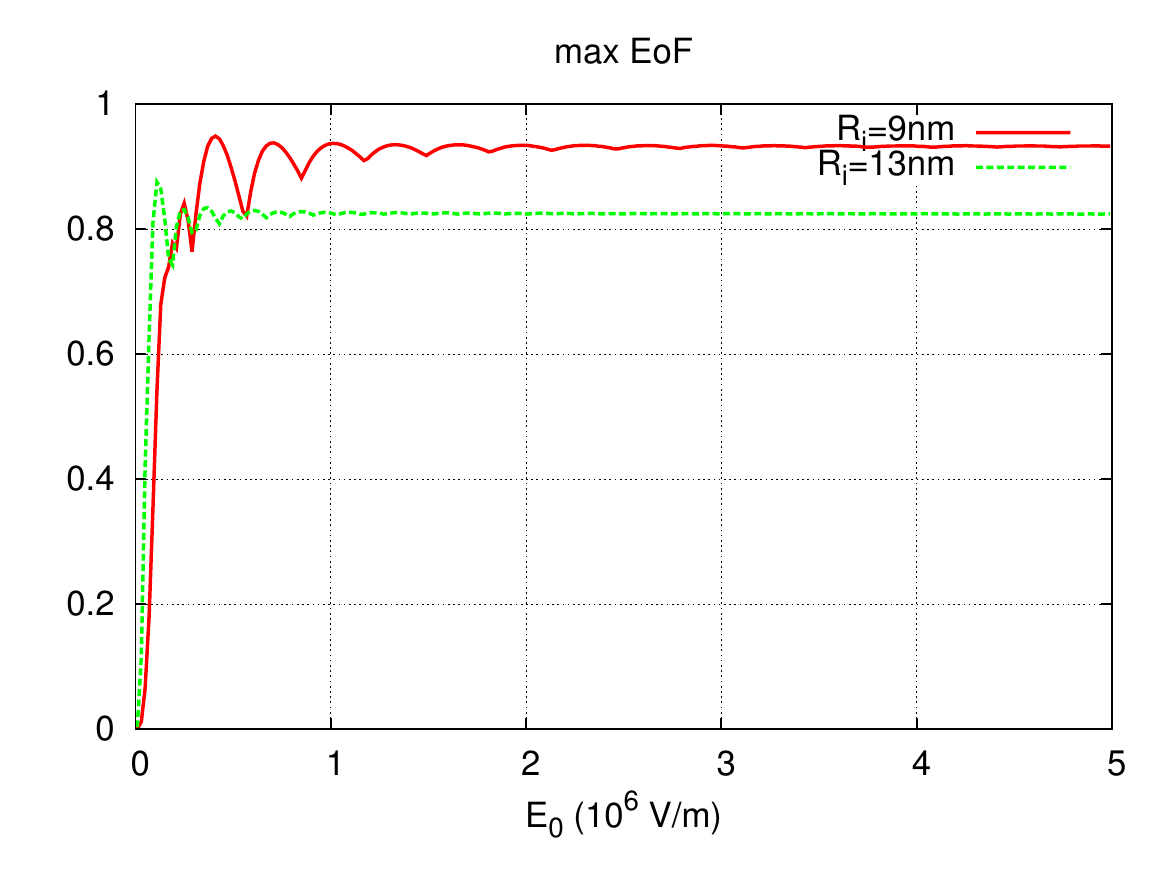}
\includegraphics[scale=0.45]{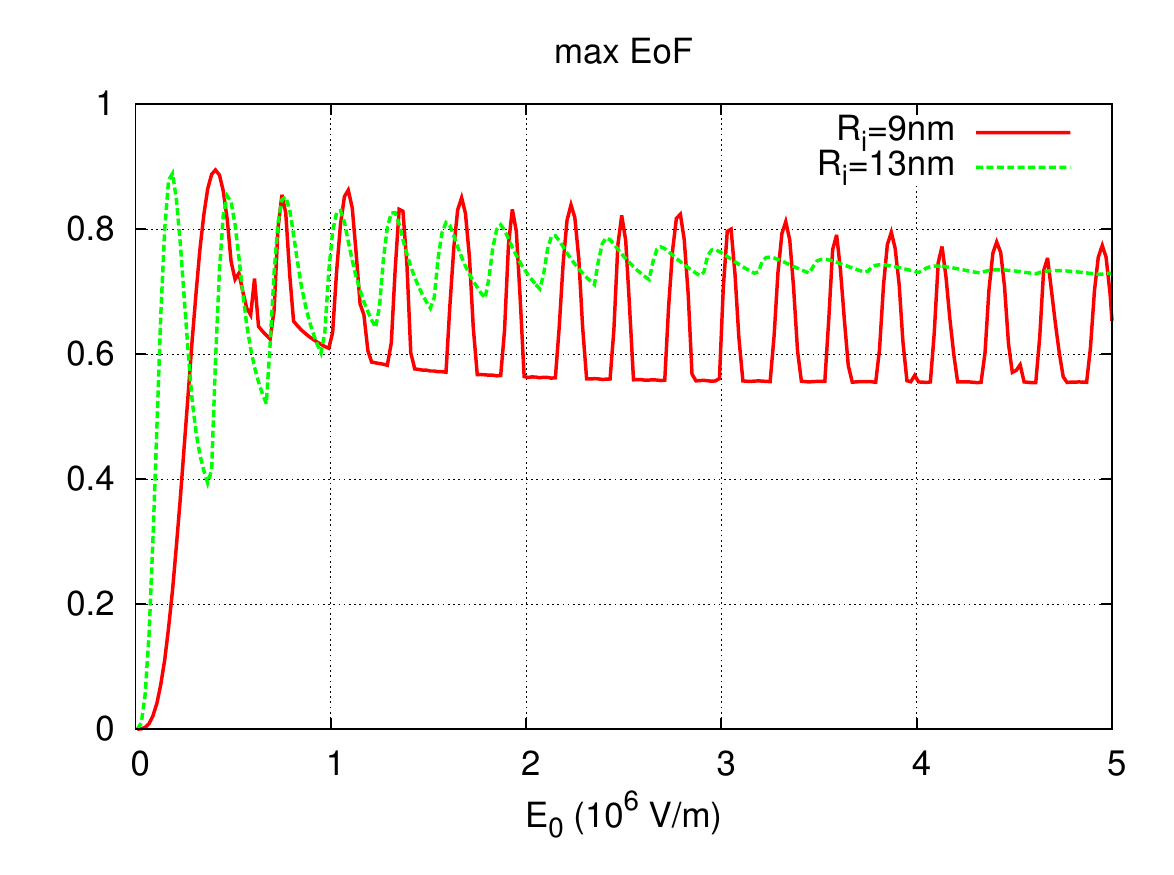}

\includegraphics[scale=0.45]{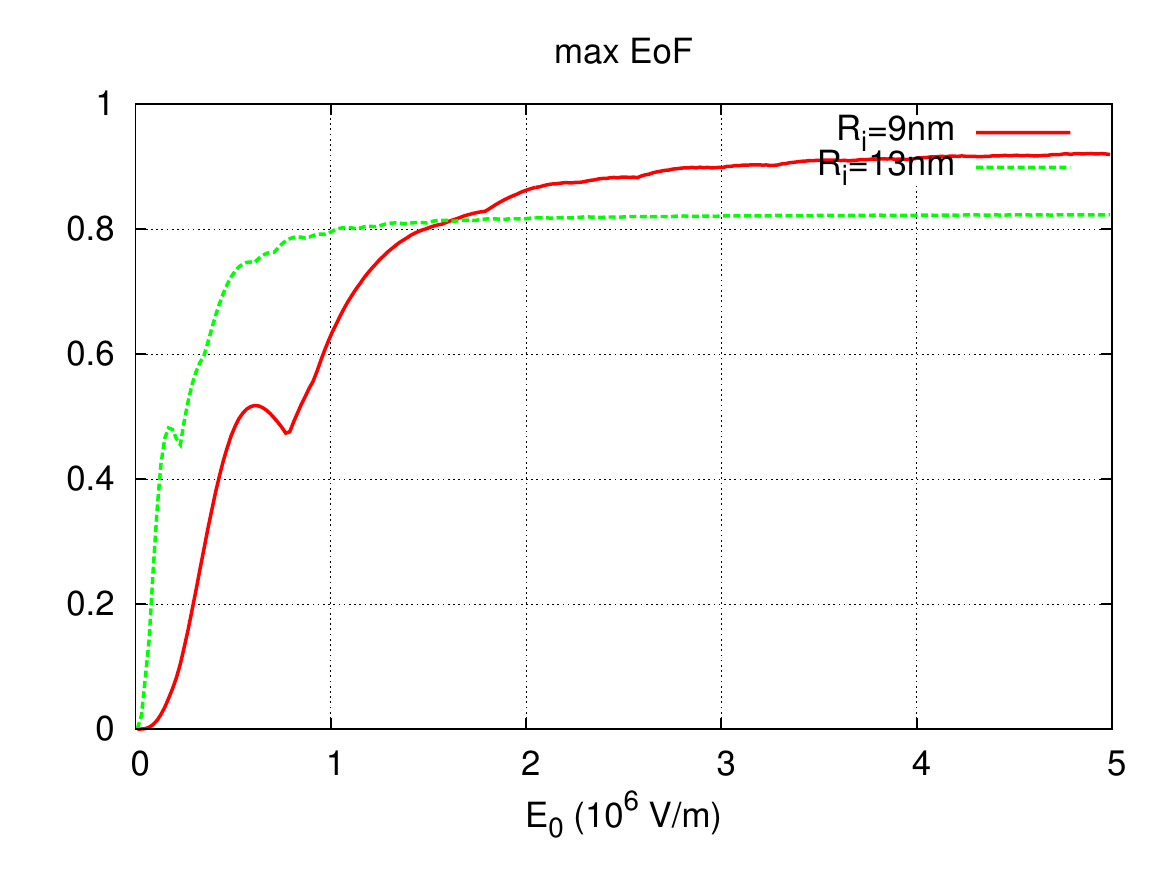}
\includegraphics[scale=0.45]{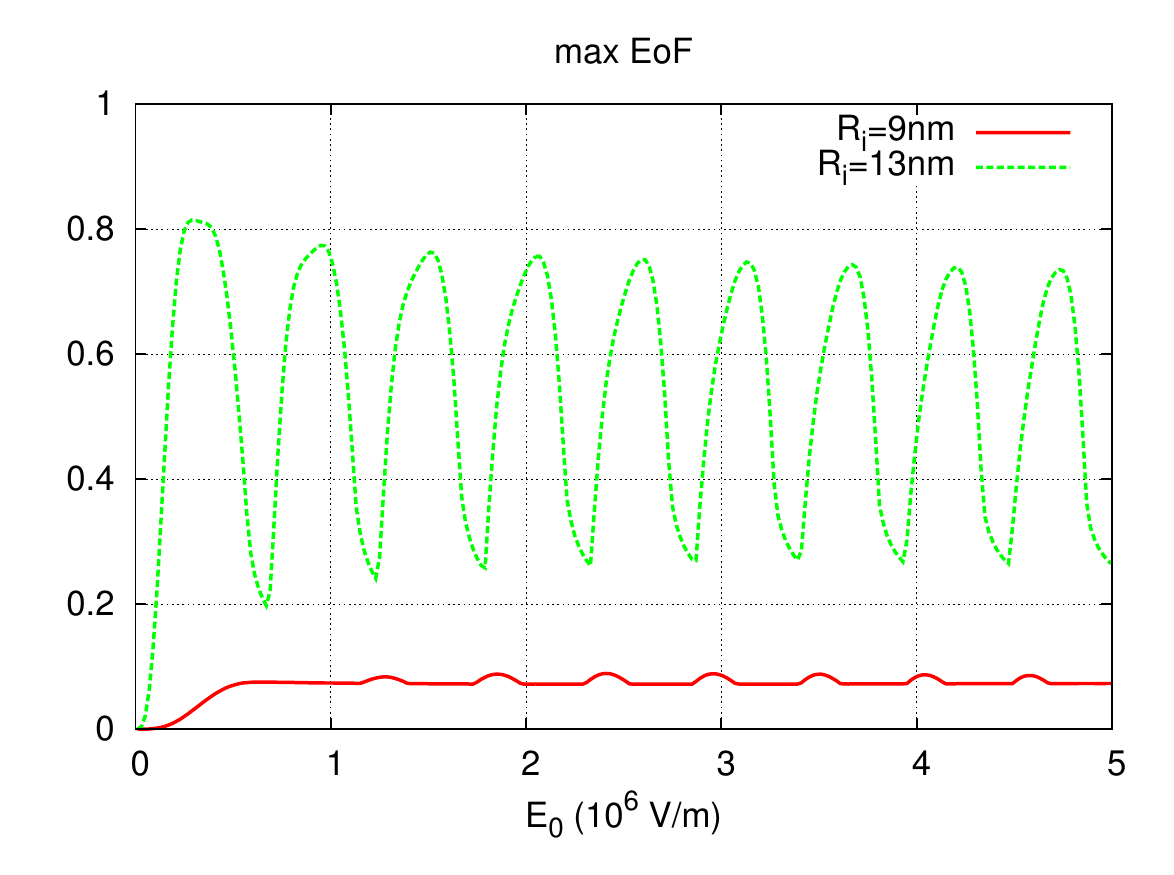}

\includegraphics[scale=0.45]{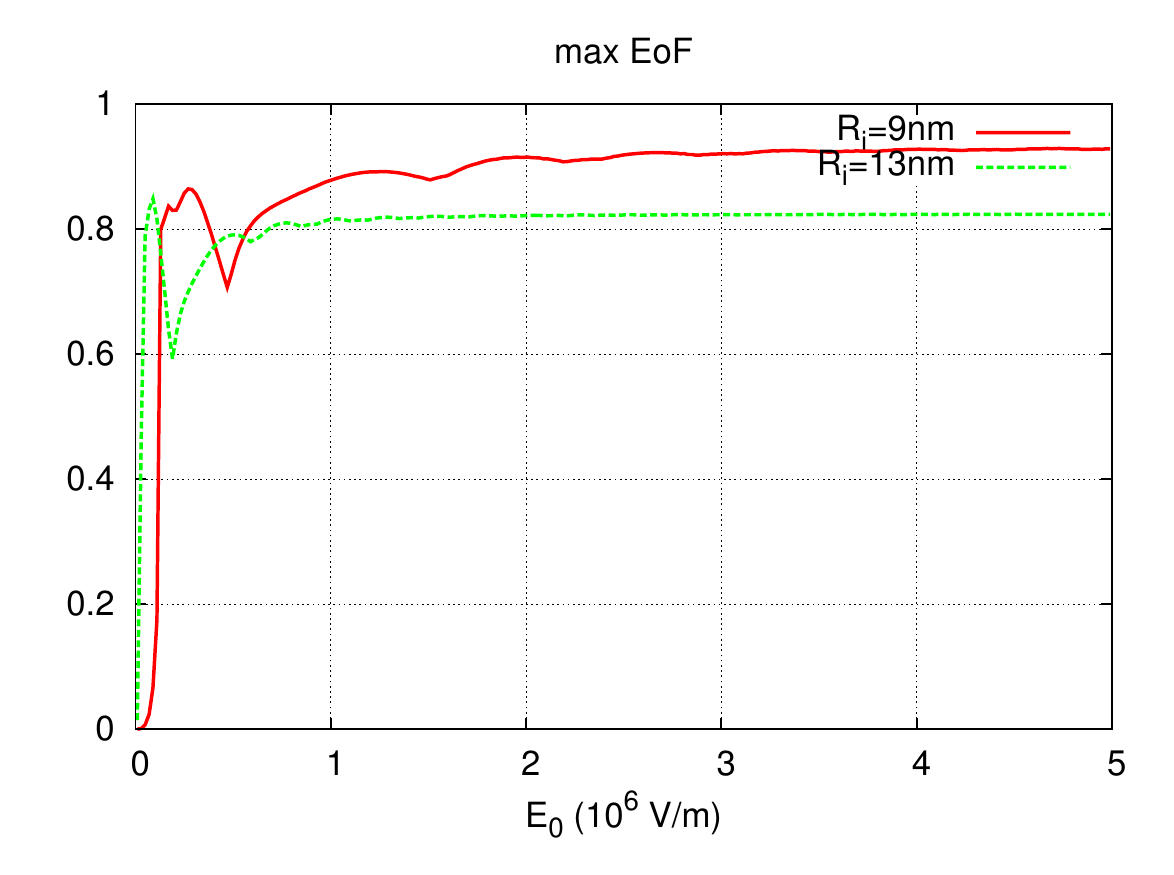}
\includegraphics[scale=0.45]{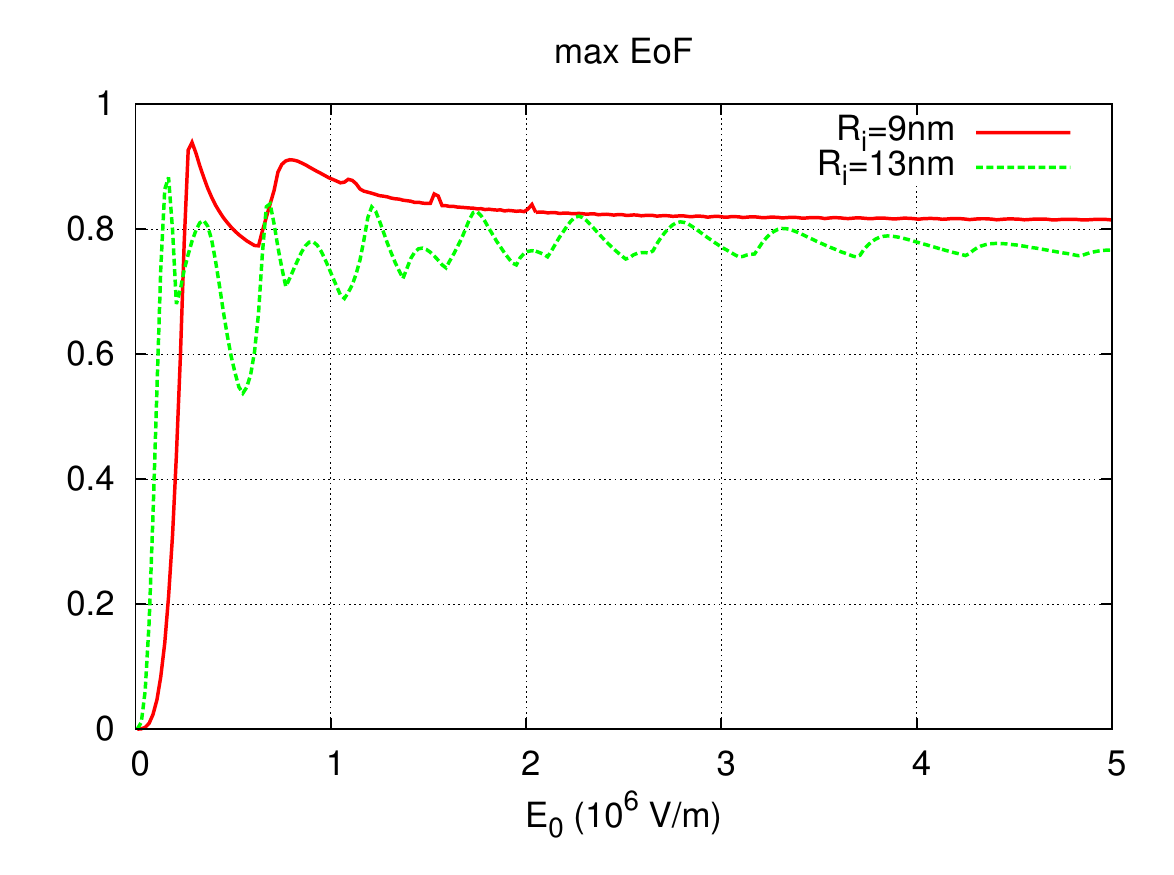}
\caption{Plots of maximum EoF as a function of $E_0$ for a hybrid system on resonance (top),
	$\omega_0 - \omega_i = -\delta$ (middle),
	$\omega_0 - \omega_i = \delta$ (bottom).
	Radius of MNP is 2 nm, the dipole moment of each SQD is \SI{2.20}{e nm} and
	the distance of each SQD from the MNP is \SI{9}{nm} (solid line) and \SI{13}{nm} (dashed).
	cw field (left) and secant pulse field (right).
	}\label{fig:eof-intensity}
\end{figure}

\begin{figure}[htp]
\centering
\includegraphics[scale=0.45]{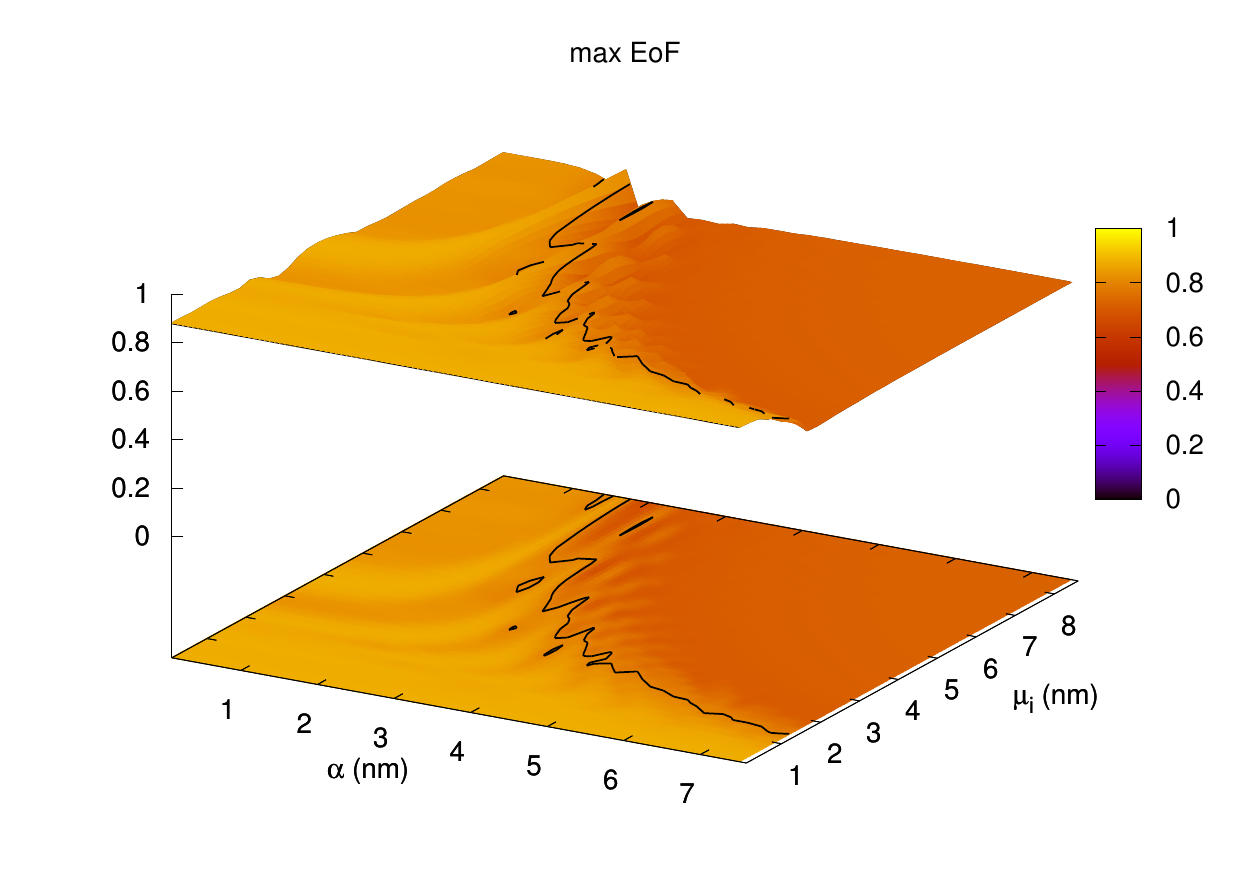}
\includegraphics[scale=0.45]{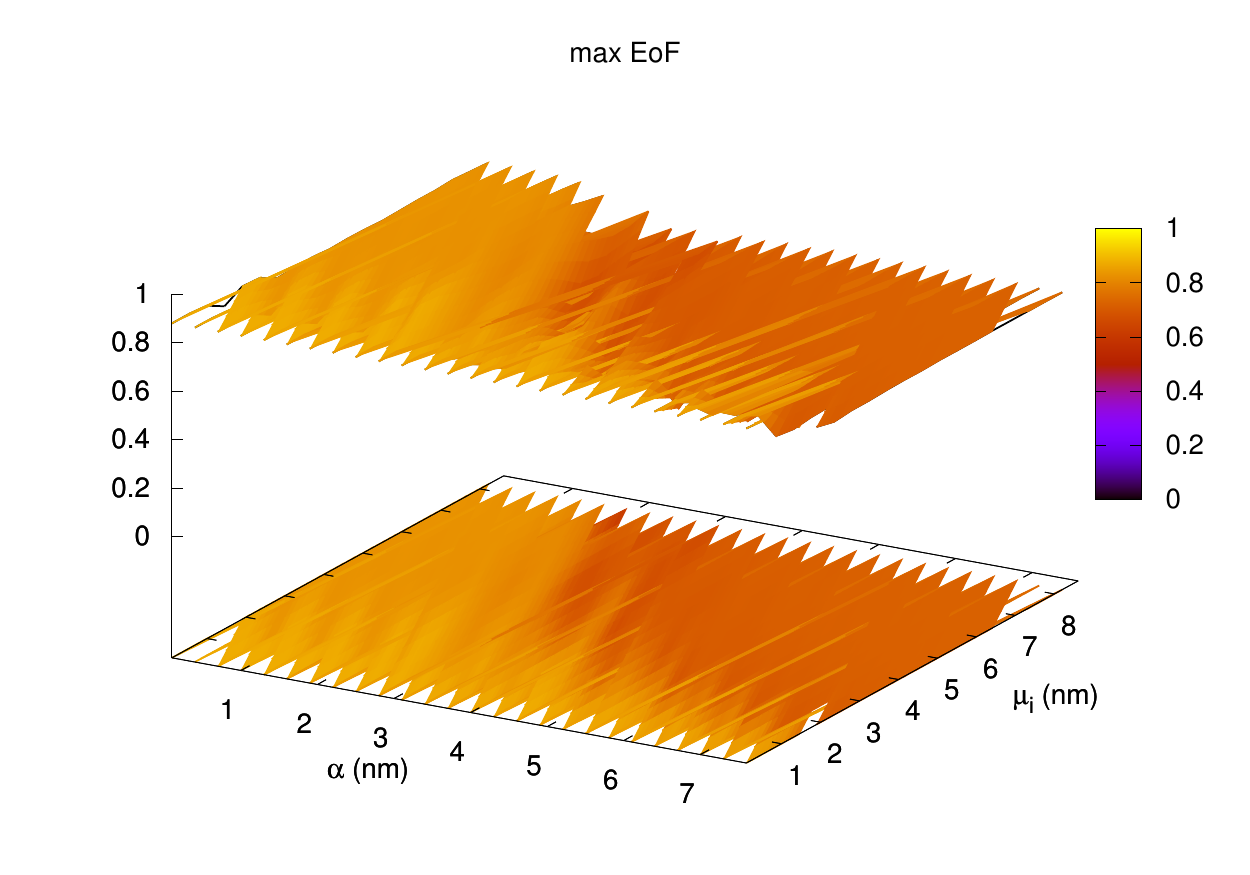}

\includegraphics[scale=0.45]{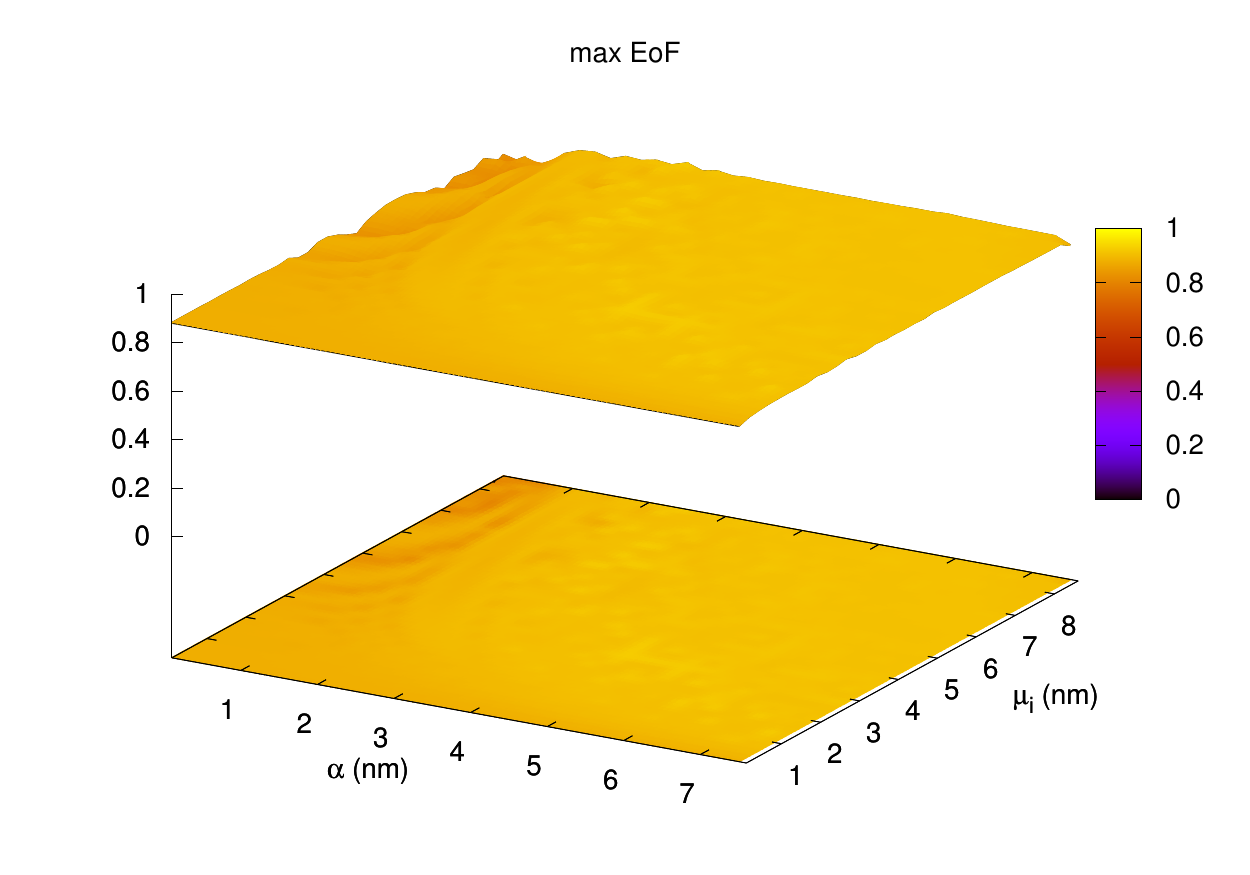}
\includegraphics[scale=0.45]{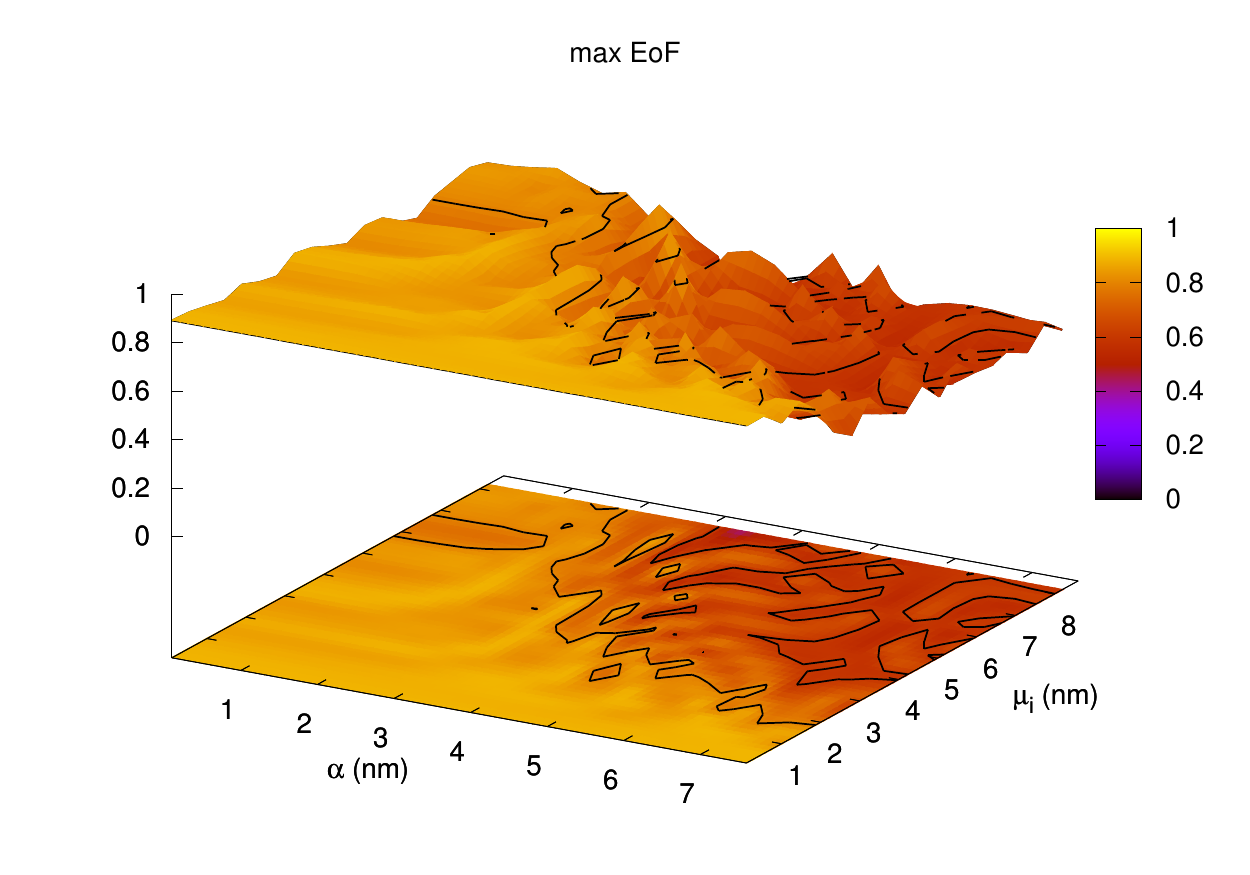}
\caption{Dipole moment ($\mu_1=\mu_2$) vs. $\alpha$ phase diagram 
	for maximum EoF achieved.
	cw field (top) and pulsed field (bottom). 
	$R_i = \SI{9}{nm}$ (left), $R_i = \SI{13}{nm}$ (right).  On the ($\mu$,$\alpha$) plane we depict the contour plot of the projection of the 3D picture.
}\label{fig:dens_alfa-miu-max-e0}
\end{figure}
\begin{figure}[htp]
\centering
\includegraphics[scale=0.45]{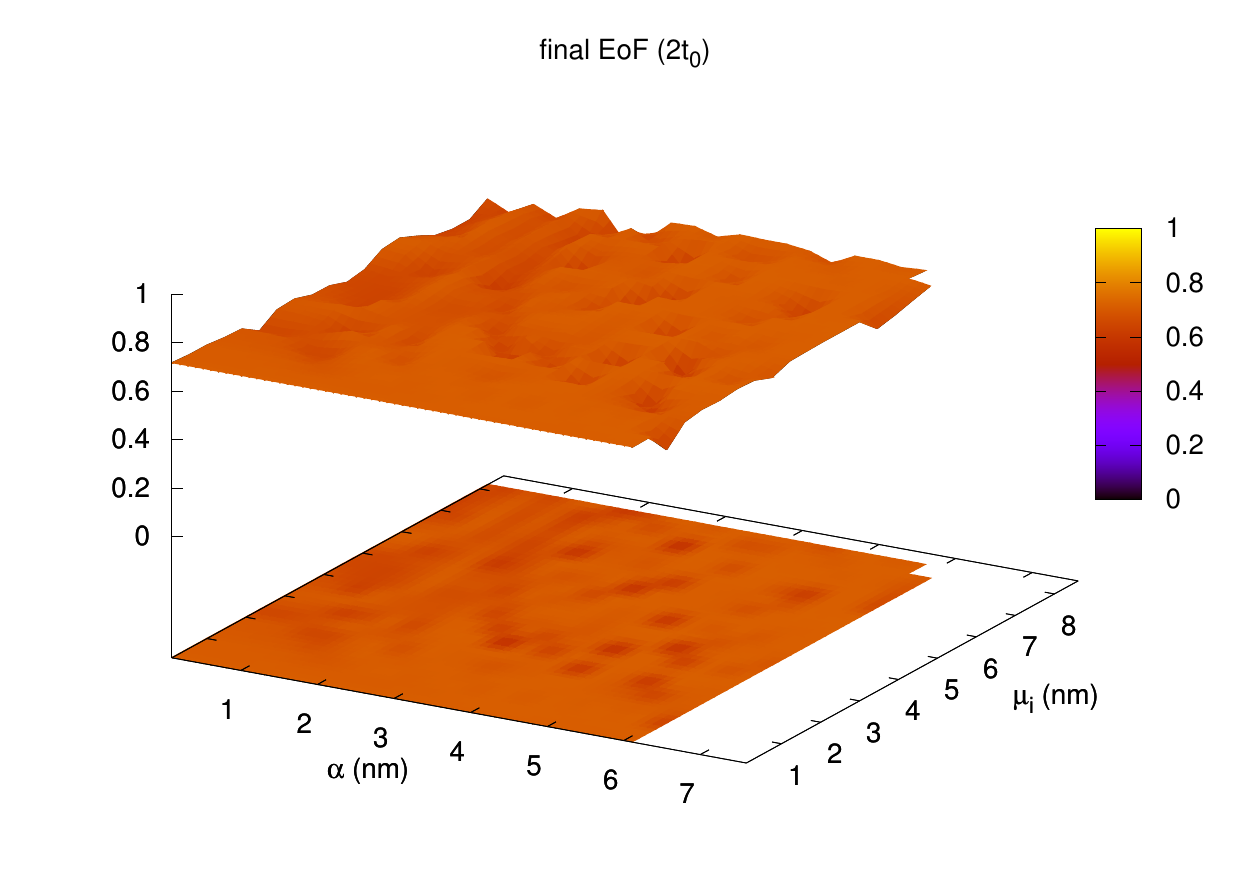}
\includegraphics[scale=0.45]{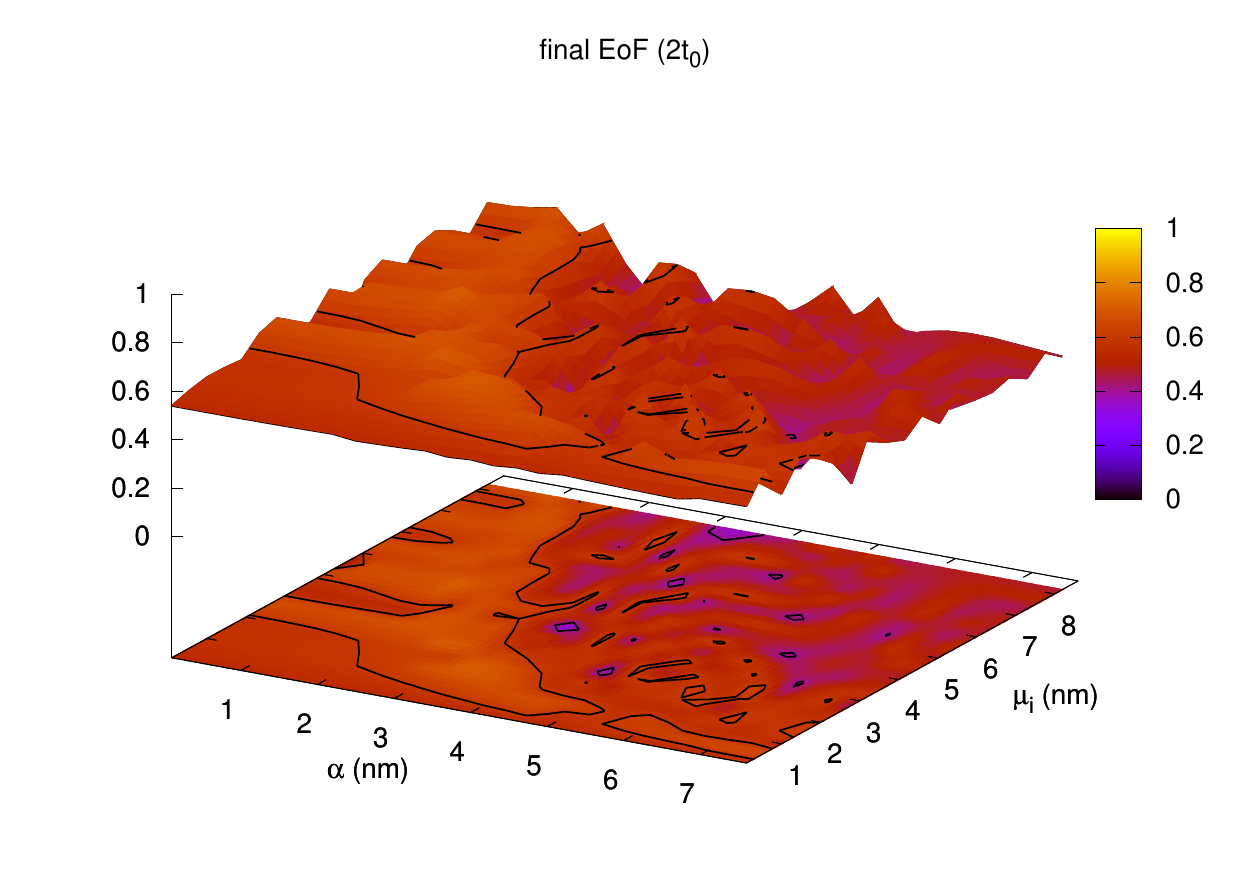}
\caption{Comparison of maximum (see previous diagram) and final EoF
	for the dipole moment ($\mu_1=\mu_2$) vs. $\alpha$ phase diagram
	for the pulsed field.
	Final EoF is measured at $2t_0$.
	$R_i = \SI{9}{nm}$ (left), $R_i = \SI{13}{nm}$ (right).
}\label{fig:dens_alfa-miu-max-e0vfinal}
\end{figure}

\begin{figure}[htp]
\centering
\includegraphics[scale=0.7]{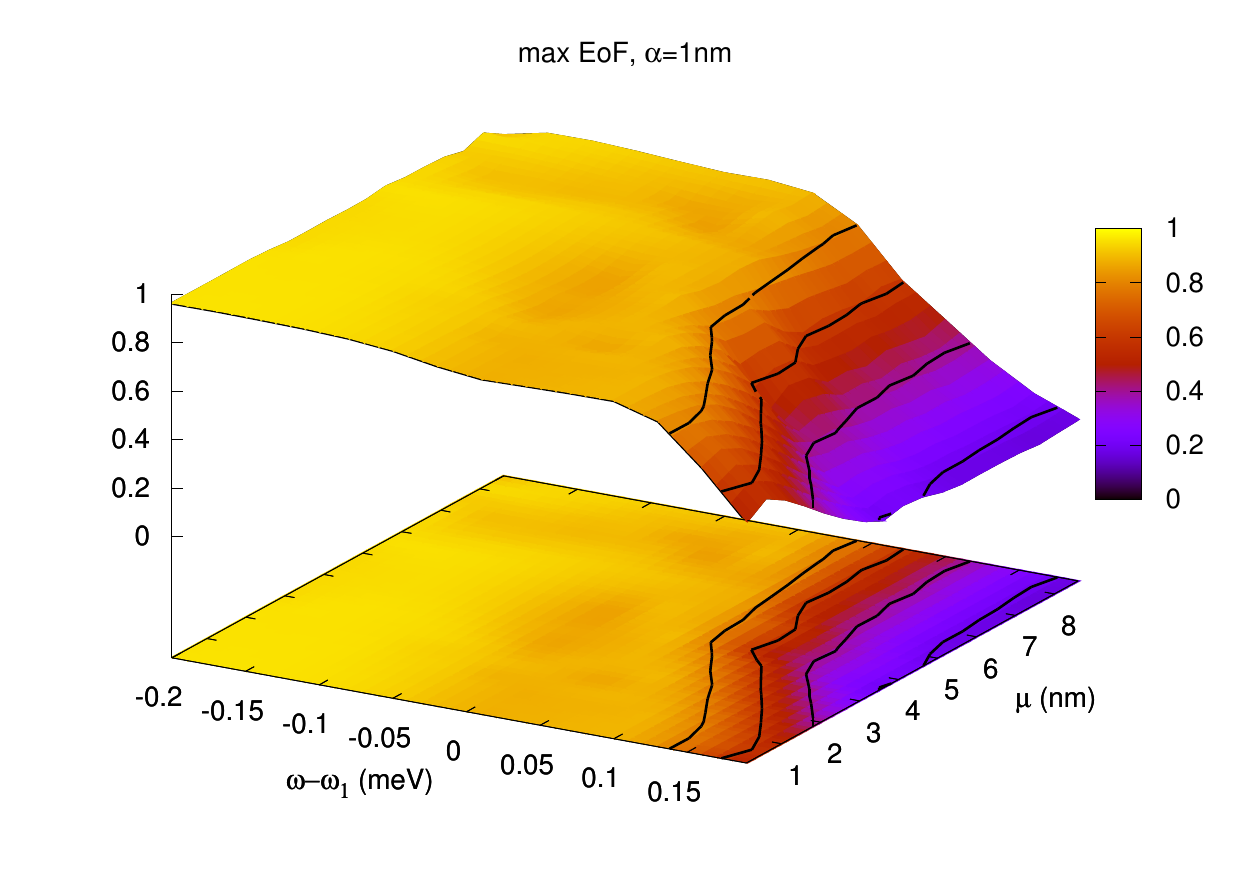}
\caption{ Dipole moment $\mu$ vs.\ detuning phase diagram 
for maximum EoF.
The parameters are the same as in the previous phase diagram.}
\label{fig:dens_Dipole-detuning}
\end{figure}

\end{document}